\documentclass[technote,10pt]{IEEEtran}
\usepackage{amsmath,amsfonts}
\usepackage{algorithmic}
\usepackage{algorithm}
\usepackage{array}
\usepackage{textcomp}
\usepackage{stfloats}
\usepackage{url}
\usepackage{verbatim}
\usepackage{float} 
\usepackage{subfig}
\usepackage{graphicx}
\usepackage{cite}
\usepackage{makecell}
\usepackage{indentfirst}
\usepackage{xcolor}
\usepackage{caption}
\usepackage{balance}

\begin{document}

\title{Pre-Equalization Aided Grant-Free Massive Access in Massive MIMO System}

\author{Yueqing Wang, Yikun Mei, Zhen Gao, Ziwei Wan, Boyu Ning, De Mi, and Sami Muhaidat

\vspace*{-3.0mm}

\thanks{The work of Zhen Gao was supported in part by Beijing Nova Program, in part by Beijing Natural Science Foundation under Grant L242011, in part by the Natural Science Foundation of China (NSFC) under and Grant 62471036 and Grant U2233216,  in part by Shandong Province Natural Science Foundation under Grant ZR2022YQ62. \emph{(Corresponding authors: Zhen Gao.)}}

\thanks{Yueqing Wang, Yikun Mei, Zhen Gao and Ziwei Wan are with the School of Information and Electronics, Beijing Institute of Technology, Beijing 100081, China (e-mails: 17710820190@163.com; meiyikun@bit.edu.cn).}

\thanks{Z. Gao is with the MIIT Key Laboratory of Complex-Field Intelligent Sensing, Beijing Institute of Technology, Beijing 100081, China, also with the Yangtze Delta Region Academy, Beijing Institute of Technology (Jiaxing), Jiaxing 314019, China, and also with the Advanced Technology Research Institute, Beijing Institute of Technology, Jinan 250307, China (e-mail: gaozhen16@bit.edu.cn)}

\thanks{Ziwei Wan are with the School of Information and Electronics, Beijing Institute of Technology, Beijing 100081, China (e-mail: ziweiwan@bit.edu.cn).}

\thanks{Boyu Ning is with the National Key Laboratory of Wireless Communications, University of Electronic Science and Technology of China, Chengdu 611731, China (e-mail: boydning@outlook.com).}

\thanks{De Mi is with School of Computing and Digital Technology,
Birmingham City University, Birmingham B55JU, UK (e-mail:
de.mi@bcu.ac.uk).}

\thanks{Sami Muhaidat is with the KU 6G Research Center, Department of Computer and Information Engineering, Khalifa University, Abu Dhabi, UAE (email: sami.muhaidat@ku.ac.ae).}
}



\IEEEpubid{}

\maketitle

\begin{abstract}
The spatial diversity and multiplexing advantages of massive multi-input-multi-output (mMIMO) can significantly improve the capacity of massive non-orthogonal multiple access (NOMA) in machine type communications. However, state-of-the-art grant-free massive NOMA schemes for mMIMO systems require accurate estimation of random access channels to perform activity detection and the following  coherent data demodulation, which suffers from excessive pilot overhead and access latency. To address this, we propose a pre-equalization aided grant-free massive access scheme for mMIMO systems, where an iterative detection scheme is conceived. Specifically, the base station (BS) firstly activates one of its antennas (i.e., beacon antenna) to  broadcast a beacon signal, which facilitates the user equipment (UEs) to perform downlink channel estimation and pre-equalize the uplink random access signal with respect to the channels associated with the beacon antenna. During the uplink transmission stage, the BS detects UEs' activity and data by using the proposed iterative detection algorithm, which consists of three modules: coarse data detection (DD), data-aided channel estimation (CE), and fine DD. In the proposed algorithm, the joint activity and DD is firstly performed based on the signals received by the beacon antenna. Subsequently, the DD is further refined by iteratively performing data-aided CE module and fine DD module using signals received by all BS antennas. Our simulation results demonstrate that the proposed scheme outperforms state-of-the-art mMIMO-based grant-free massive NOMA schemes with the same access latency. Simulation codes are provided to reproduce the results in this article: https://github.com/owenwang517/tvt-2025.
\end{abstract}

\IEEEpeerreviewmaketitle
\section{Introduction}

The advent of next-generation Internet-of-Things (IoT) is envisaged to enable a ubiquitous connectivity of massive user equipment (UEs) \cite{Gao2024CSGFMA}. Recently, researchers have proposed to integrate grant-free random access protocols with non-orthogonal multiple access (NOMA) techniques \cite{Chen2021MA5G}. In such schemes, UEs leverage non-orthogonal resources to transmit data to the base station (BS) simultaneously, rather than the complicated four-step handshaking process required by the conventional grant-based random access schemes. This approach can effectively increase the upper limit of the system's access capacity and reduce the access latency. 

Meanwhile, compared with single-antenna BS, the diversity achieved by massive multiple-input-multiple-output (mMIMO) can further enhance the capacity and detection performance of massive machine-type communication (mMTC) systems. Consequently, mMIMO-based grant-free massive access schemes have attracted significant attention. Most existing works consider that UEs adopt a ``preamble+data" transmission frame structure \cite{Liu2018massiveconnectivityMIMO, Ke2020CSADCEMIMO, Ke2021cellfree, Chen2022JADCE, Zhou2023NOMAOTFS,Zheng2024DLADCEIRS, Shen2022MIMOOTFS, Shao2022lowrank}. At the receiver, the BS firstly performs joint activity detection and channel estimation (JADCE) based on the received preamble, and then perform coherent data detection using the estimated channel state information (CSI). The JADCE problem is usually solved by the compressed sensing (CS) reconstruction algorithm that leverages the sparsity of UEs' activity pattern and the sparsity of CSI in the virtual angular domain. The data detection can be performed with message passing algorithms \cite{Ge2021OTFSNOMA}. However, estimating the UEs' activity relies on the accurate estimation of the high-dimensional mMIMO CSI matrix, leading to high preamble overhead and access latency. On the other hand, assuming that the CSI of all UEs are available at the BS, some literatures focus on the joint activity and data detection (JADD) problem in the uplink random access \cite{Wei2017JADD}. To avoid this impractical assumption on CSI acquirement, \cite{Mei2022Preeq} proposed a beacon-aided grant-free random access scheme for solving JADD problem, where the uplink access signal is pre-equalized by exploiting the uplink/downlink channel reciprocity. However, \cite{Mei2022Preeq} only considers the single antenna BS configurations.

Our primary contribution lies in extending the pre-equalization aided grant-free access scheme to mMIMO systems. This framework presents distinct challenges. First, the pre-equalization methods should be more carefully investigated, as previous work only consider idealized pre-equalization method. Second, it is more challenging to achieve detection at the mMIMO receiver. As the mMIMO BS has multiple antennas, it is impossible to estimate and pre-equalize the whole mMIMO channel at the single-antenna users. Meanwhile, with the same resource overhead as \cite{Mei2022Preeq}, the users can only obtain the CSI corresponding to one of the BS’s antennas, while the majority of the whole CSI remains unknown. This limitation turns the linear estimation problem of joint AD and DD into a challenging semi-blind detection problem including AD, channel estimation (CE) and DD. In response to these challenges, we have made contributions and innovations in both pre-equalization method design and algorithm development as follows:

\begin{itemize}
    \item \textbf{Pre-equalization Method}: We propose a more robust pre-equalization method with nulling operation to suppress the magnitude of pre-equalization factors, which restricts the peak power of the transmitter in a reasonable range.
    \item \textbf{Detection Algorithm Design}: To tackle the semi-blind detection problem of joint AD, CE and DD, we propose an iterative detection algorithm by transforming this complex problem into three standard linear models. Accordingly, our proposed algorithm consists of three modules, including: a) AD and Coarse DD, b) CE, and c) Fine DD, respectively. This staged approach allows us to leverage well-established techniques for each sub-problem. Besides, the iteration between data-aided CE and refined DD can further improve the detection performance.
\end{itemize}

Simulation results show that the DD performance of the proposed scheme outperforms its counterpart tailored for single-antenna configurations in \cite{Mei2022Preeq}, thanks to the diversity advantage of mMIMO. Besides, the activity detection (AD) and DD performance of the proposed scheme surpasses the state-of-the-art schemes adopting the ``pilot+data" transmission frame structure with JADCE and coherent DD.

\textit{Notations:} Tensors, matrices, and vectors are represented in the form of $\boldsymbol{ \mathcal{H} }, \mathbf{ H }$, and $\mathbf{ h }$, respectively. For a third-order tensor $\boldsymbol{ \mathcal{H} }\in \mathbb{C}^{ M \times N \times K } $, $\boldsymbol{ \mathcal{H} }_{ m,:,: },\boldsymbol{ \mathcal{H} }_{ :,n,: },\boldsymbol{ \mathcal{H} }_{ :,:,k } $ denote slices and $\boldsymbol{ \mathcal{H} }_{ :,n,k }, \boldsymbol{ \mathcal{H} }_{ m,:,k }, \boldsymbol{ \mathcal{H} }_{ m,n,: } $ denote fibers of the tensor $\boldsymbol{ \mathcal{H} }$. $\left[ \mathbf{ h } \right]_{ n }$ represents the $n$-th element of vector $\mathbf{ h }$. $[\mathbf{ H }]_{m,n}$ represents the $(m,n)$-th element of matrix $\mathbf{ H }$. Let $\Omega$ denote a set, then $|\Omega|_{c}$ denotes the cardinality of $\Omega$. $[\mathbf{H}]_{\Omega,:}$ denotes the sub-matrix that stacks row vectors of matrix $\mathbf{ H }$ indexed by $\Omega$. $\mathbf{ 1 }_{ M \times N } \in \mathbb{C}^{ M \times N } $ denotes an all-ones matrix. $\mathbf{ I }_{M\times M} \in \mathbf{C}^{ M \times M } $ refers to an identity matrix. $(\cdot)^{\mathrm{T}} ,(\cdot)^{\mathrm{H}} ,(\cdot)^*$ denotes the transpose, Hermitian, and conjugation operation, respectively. $\circ$ refers to the Hadamard product.
\IEEEpubidadjcol

\section{System Model}

\begin{figure}
    \centering
    \includegraphics[width=1.0\linewidth]{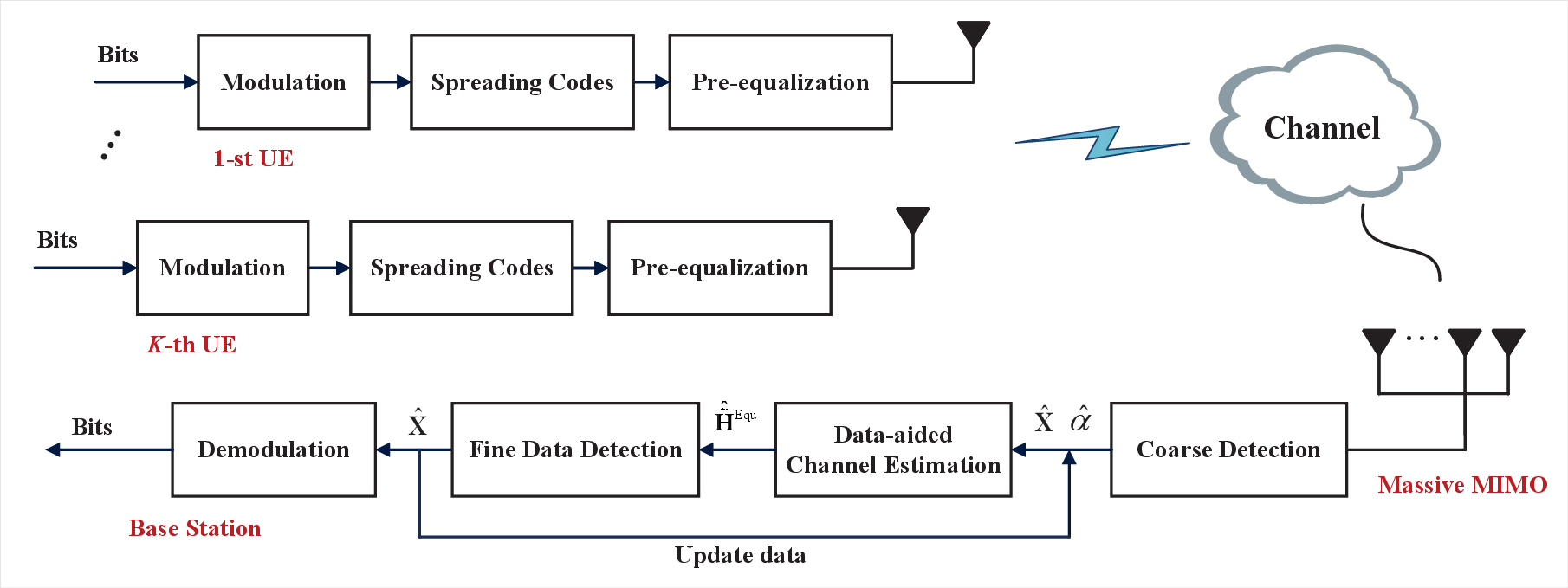}
    \caption {The proposed pre-equalization aided grant-free massive access scheme.}
    \label{fig:system_model}
\end{figure}

We consider a typical massive access scenario, where a time division duplexing mMIMO BS is equipped with a uniform linear array (ULA) of $N$ antennas and services $K$ single-antenna UEs. We employ OFDM to combat frequency-selective fading channels.

\vspace{5pt}
We denote $\breve{\boldsymbol{\mathcal{H}}}\in \mathbb{C}^{ N \times M \times K }$ and $\boldsymbol{\mathcal{H}}\in \mathbb{C}^{ N \times M \times K }$ as all $K$ UEs' large-scale fading channel tensor and small-scale fading channel tensor in the spatial-frequency domain (i.e., $N$ BS antennas $\times$ $M$ subcarriers), respectively. For example, $\boldsymbol{\mathcal{H}}_{ n,m,k }$ represents the small-scale channel fading factor from the $k$-th UE to the $n$-th BS antenna on the $m$-th subcarrier. For the $m$-th subcarrier, the channel of the $k$-th UE is modeled as $\breve{\boldsymbol{\mathcal{H}}}_{ :,m,k } =\sqrt{g_{ k } }\boldsymbol{\mathcal{H}}_{ :,m,k } \in \mathbb{C}^{N\times 1} $, where $g_{ k }$ denotes large-scale channel fading factor. BS is elevated high and surrounded by few scatterers, while UEs are usually located in a rich-scattering environment \cite{Gao2015SparsityMIMO}. Therefore, the classic one-ring model is adopted to model the wireless channels. Hence, the small-scale fading factor $\boldsymbol{ \mathcal{ H } }_{ :,m,k }$ can be modeled as follows
\vspace{-3pt}
\begin{equation}
	\label{Eq:MIMOChannel}
	\boldsymbol{\mathcal{H}}_{:,m,k}=\sum_{p=1}^P{\beta _{k,p}\boldsymbol{\alpha }_R\left( \phi _{k,p} \right) e^{-j2\pi \tau _{k,p}\left[ -\frac{B_s}{2}+\frac{\left( m-1 \right)}{M}B_s \right]}},
\end{equation}
where $P$ denotes the number of paths, $\phi_{ k,p } $ and $\tau_{ k,p } $ denote the angle of arrival and the delay of the $p$-th path, respectively. The steering vector is given by $\boldsymbol{\alpha }_R \left( \phi _{k,p} \right) =\left[ 1,e^{-j\frac{2\pi}{\lambda}d\sin \left( \phi _{k,p} \right)},\cdots ,e^{-j\frac{2\pi}{\lambda}\left( N-1 \right) d\sin \left( \phi _{k,p} \right)} \right] ^T \in \mathbb{C}^{N \times 1}$. $B_{s}$ is the bandwidth, $\lambda$ represents the wave length, and $d = \frac{\lambda}{2}$ is the antenna spacing.

The proposed pre-equalization aided grant-free massive access scheme includes two primary stages. At the first stage, the BS randomly activates one antenna (named beacon antenna) to periodically broadcast the beacon to all UEs. After receiving the beacon, UEs conduct synchronization and estimate the downlink channels. Subsequently, at the second stage, UEs transmit uplink random access signals to the BS with power controlling and pre-equalization by exploiting the uplink/downlink channel reciprocity. A beacon broadcast period precedes each frame dedicated to uplink transmission. Throughout one frame for uplink transmission, each active UE sends $T$ OFDM symbols to the BS. Each UE is assigned an unique, albeit non-orthogonal, spreading code on $M$ subcarriers. Let $\mathbf{ s }_{ k } \in \mathbb{C}^{M\times 1} $ denote the spreading codes of the $k$-th UE and the modulus of all elements of $\mathbf{ s }_{ k }$ are equal to $1$. Upon receiving the superimposed signal from active UEs within a frame, the BS performs active UEs identification and iterative DD. The overall uplink procedure in shown in Fig. \ref{fig:system_model}.
 
\section{Preprocessing at the UEs}

Upon receiving the beacon signal, UEs calculate the pre-equalization factors according to the downlink CE result. Let $\eta$ represent the index of the beacon antenna, whose channel will be respectively pre-equalized by different UEs. Then, we use $\hat{\boldsymbol{ \mathcal{H} }}_{\eta,m,k}$ to represent the estimation of the small-scale fading factor estimated by the $k$-th UE with respect to the $m$-th subcarrier based on the received beacon. Here, we suppose the CE at UEs' side is sufficiently accurate, i.e., $\hat{\boldsymbol{ \mathcal{H} }}_{\eta,m,k} = \boldsymbol{ \mathcal{H} }_{\eta,m,k}$. The pre-equalization factor of the $k$-th UE with respect to the $m$-th subcarrier is represented as $\theta _{ m,k } $. 
We apply the pre-equalization with nulling operation in equation \ref{Eq:Pre-eq}. \footnote{ The reason is that when the channel experiences severe attenuation, the magnitude of pre-equalization factors can become excessively large, leading to a degradation in overall system performance. This degradation is caused by the saturation of the transmitter power amplifier and reduced SNR at the beacon antenna.}
\begin{equation}
	\label{Eq:Pre-eq}
	\begin{cases}
		\theta _{m,k}=1/\hat{\boldsymbol{ \mathcal{H} }}_{\eta,m,k}, &	\mathrm{if} \,\,|\hat{\boldsymbol{ \mathcal{H} }}_{\eta,m,k}|\ge h_0,\\
		\theta _{m,k}=0, &		\mathrm{if}\,\,|\hat{\boldsymbol{ \mathcal{H} }}_{\eta,m,k}| < h_0,\\
	\end{cases}
\end{equation}
where $h_{ 0 }$ is a predefined threshold.  Furthermore, the pre-equalization factors with respect to all subcarriers at the $k$-th UE can be expressed by a vector $\boldsymbol{ \theta }_{ k } = [\theta_{1,k},\theta_{2,k},\cdots,\theta_{M,k}]^{\mathrm{T}}  \in \mathbb{C}^{ M\times 1 } $. The average power of pre-equalization factor is denoted as $p_{ e } $, i.e., $\mathbb{E}[\left| \theta_{m,k} \right|^{ 2 }   ]= p_{ e } $.

At the UE side, the data is multiplied with spreading codes, pre-equalization factors, and power controlling factors. Hence, the signal transmitted by the $k$-th UE can be expressed as $\sqrt{p_{ k } }\left( \boldsymbol{ \theta }_{ k }  \circ \mathbf{ s }_{ k }  \right)\alpha_{ k } x_{ k,t } \in \mathbb{C}^{ M\times 1 } $, where $x_{ k,t } $ denotes the data symbol transmitted by the $k$-th UE in the $t$-th time slot. The value of $x_{ k,t } $ belongs to a modulation constellation set $\Omega = \{a_{ 1 }, a_{ 2 }, \cdots, a_{ L } \}$, where $L$ denotes the order of modulation. When the $k$-th UE is inactive during one frame, $x_{ k,t } $ equals $0$. $\alpha_{ k }\in\left\{0,1 \right\}  $ represents the activity indicator of the $k$-th UE. If the $k$-th UE is active, $\alpha_{ k }=1$. $p_{ k } =1/g_{ k }$ denotes the power controlling factor of the $k$-th UE, which is inversely proportional to the large-scale fading factor $ g_{ k } $. Therefore, the transmit power of the $k$-th UE is expressed as $\rho_{ k }  = p_{ k }  p_{ e } = p_{ e }/g_{ k }  $. 

After the above preprocessing, signal received by the BS within one frame can be represented by a third-order tensor $\boldsymbol{\mathcal{Y}} \in \mathbb{C}^{ N \times M \times T } $. Therefore, the signal received by the $n$-th BS antenna in the $t$-th time slot $\boldsymbol{ \mathcal{Y} }_{ n,:,t }^{\mathrm{T}}  \in \mathbb{C}^{ M \times 1 } $ can be expressed as follows
\begin{equation}
	\begin{aligned}
		\label{Eq:model_vec}
		\boldsymbol{ \mathcal{Y} }_{ n,:,t }^{\mathrm{T}}  &= \sum_{ k=1 }^{ K } \sqrt{p_{ k } }\left( \breve{\boldsymbol{ \mathcal{H} }}_{ n,:,k }^{\mathrm{T}}  \circ \boldsymbol{ \theta }_{ k } \circ \mathbf{ s }_{ k }   \right) \alpha_{ k }  x_{ k,t }  + \boldsymbol{ \mathcal{W} }_{ n,:,t }^{\mathrm{T}}  \\
		&= \sum_{ k=1 }^{ K } \left( \boldsymbol{ \mathcal{H} }_{ n,:,k }^{\mathrm{T}}   \circ \boldsymbol{ \theta }_{ k } \circ \mathbf{ s }_{ k } \right) \alpha_{ k } x_{ k,t } +  \boldsymbol{ \mathcal{W} }_{ n,:,t }^{\mathrm{T}} ,
	\end{aligned}
\end{equation}
where $\breve{\boldsymbol{ \mathcal{H} }}_{ n,;,k }^{\mathrm{T}}  = \sqrt{g_{ k }}\boldsymbol{ \mathcal{H} }_{ n,:,k }^{\mathrm{T}} \in \mathbb{C}^{ M \times 1 }$ represents the large-scale channel fading from the $k$-th UE to the $n$-th antenna of BS. Besides, the additive Gaussian noise is represented as a third-order tensor $\boldsymbol{ \mathcal{W} } \in \mathbb{C}^{ N \times M \times T }  $ and $\boldsymbol{ \mathcal{W} }_{ n, :, t }^{\mathrm{T}}$ obeys the distribution $\mathcal{CN}(\mathbf{ 0 },\sigma^{ 2 } \mathbf{ I }_{ M \times M } )$. 

Furthermore, the data received by the $n$-th antenna at the BS in one frame, i.e. $\boldsymbol{ \mathcal{Y} }_{ n,:,: } \in \mathbb{C}^{M \times T}$ can be expressed as follows
\begin{equation}
	\label{Eq:model_mtx}
	\boldsymbol{ \mathcal{Y} }_{n,:,:}=\left( \boldsymbol{ \mathcal{H} }_{n,:,:} \circ \boldsymbol{\Theta } \circ \mathbf{S} \right) \mathbf{X} + \boldsymbol{ \mathcal{W} }_{n,:,:}, \quad n=1,2,\cdots,N,
\end{equation}
where $\boldsymbol{\Theta}= [\boldsymbol{ \theta}_{1} , \boldsymbol{ \theta }_{ 2 }, \cdots,\boldsymbol{ \theta }_{ K }  ] \in \mathbb{C}^{M \times K}$ and $\mathbf{S} = [\mathbf{s}_1,\mathbf{s}_2,\cdots,\mathbf{s}_K] \in \mathbb{C}^{M \times K}$. The data symbol along with the activity of UEs are represented by matrix $\mathbf{X} \in \mathbb{C}^{K \times T}$, where $[\mathbf{ X }]_{k,t}=\alpha_{ k } x_{k,t}$.

\section{Proposed Receiver Detection Scheme at the BS}

The proposed receiver detection scheme primarily consists of coarse DD module, data-aided CE module, and fine DD module. In the coarse DD module, due to the pre-equalization procedure in equation (\ref{Eq:Pre-eq}), the BS can coarsely detect UEs' activity and data symbols according to the signal received from beacon antenna without the knowledge CSI. Afterwards, regarding the estimated data as pilot, the BS can conduct data-aided CE using the signals received by all antennas, i.e., data-aided CE module. In the fine DD module, with the CSI estimated by the data-aided CE module, the BS performs a refined DD using the signals received by all antennas. Finally, we iteratively perform data-aided CE and fine DD to enhance the accuracy of DD.

\subsection{The First Module: Coarse Data Detection}

We assume that the small-scale fading factors of the beacon antenna $\boldsymbol{ \mathcal{H} }_{ \eta,:,: } $ is completely equalized by $\boldsymbol{ \Theta }$, i.e., $\boldsymbol{ \mathcal{H} }_{ \eta,:,: } \circ \boldsymbol{ \Theta } = \mathbf{ 1 }$. Hence, the signal received by beacon $\boldsymbol{ \mathcal{Y} }_{ \eta,:,: } $ can be expressed as follows
\begin{equation}
	\label{Eq:JADD_MMV}
	\boldsymbol{ \mathcal{Y} }_{\eta,:,:} =\left( \boldsymbol{ \mathcal{H} }_{\eta,:,:} \circ \boldsymbol{\Theta } \circ \mathbf{S} \right) \mathbf{X} + \boldsymbol{ \mathcal{W} }_{\eta,:,:} \\
	= \mathbf{S} \mathbf{X} + \boldsymbol{ \mathcal{W} }_{\eta,:,:}.
\end{equation} 

Considering that only a small fraction of UEs are active in single frame, every column vector of $\mathbf{X}$ is sparse. Moreover, all the column vectors of $\mathbf{X}$ share common support set due to the invariant UEs' activity in one frame. Therefore, the problem of activity and DD can be modeled as a multiple measurement vectors (MMV) sparse signal recovery problem. To handle the problem, we adopt the approximate message passing (AMP) algorithm.
 
We adopt the spike-and-slab distribution that is commonly used in sparse recovery problems \cite{Wei2017JADD}, \cite{Mei2022Preeq} as \textit{a priori} distribution of $x_{k,t}$ to capture the sparse nature of the uplink traffic, i.e.,
\begin{equation}
	\label{Eq:prior}
	p_{0}\left(x_{k, t}\right)=\left(1-\gamma_{k, t}\right) \delta\left(x_{k, t}\right) \\
	+\frac{\gamma_{k, t}}{L} \sum_{l=1}^{L} \delta\left(x_{k, t}-a_{l}\right),
\end{equation}
where the sparsity ratio $\gamma_{k,t}$ refers to the probability that variable $x_{k,t}$ is nonzero and $\delta(\cdot)$ refers to Dirac delta function.

The update rule of the variance and mean at factor nodes in the $i$-th iteration are as follows, and $\hat{x}_{k, t}^{i}$ denotes the posterior mean of $x_{k,t}$.
\begin{equation}
	\label{Eq:update_factor}
	\begin{aligned}
		& V_{m, t}^{i}=\sum_{k}\left|s_{m, k}\right|^{2} v_{k, t}^{i}, \\
		& Z_{m, t}^{i}=\sum_{k} s_{m, k} \hat{x}_{k, t}^{i}-{V_{m, t}^{i}}/\left[ {\sigma^2+V_{m, t}^{i-1}\left(\boldsymbol{ \mathcal{Y} }_{\eta, m, t}-Z_{m, t}^{i-1}\right)}\right].
	\end{aligned}
\end{equation}

The update rule of the variance and mean of variable nodes in the $i$-th iteration are as follows 
\begin{equation}
	\label{Eq:update_variable}
	\begin{aligned}
		D_{k, t}^{i} & =\left[\sum_{m} {\left|s_{m, k}\right|^{2}}/{(\sigma^2+V_{m, t}^{i})}\right]^{-1}, \\
		C_{k, t}^{i} & =\hat{x}_{k, t}^{i}+D_{k, t}^{i} \sum_{m} \frac{s_{m, k}^{*}\left(\boldsymbol{ \mathcal{Y} }_{\eta, m, t}-Z_{m, t}^{i}\right)}{\sigma^2+V_{m, t}^{i}}.
	\end{aligned}
\end{equation}

According to the sum-product algorithm and the AMP algorithm \cite{Donoho2010AMP}, the marginal \textit{a posteriori} distribution of $x_{k,t}$ is the product of the prior distribution and the message to the variable node shown as equation (\ref{eq:post_margin}). The result of the product operation is another discrete spike-and-slab distribution. Consequently, we can express the marginal posterior distribution as a discrete spike-and-slab distribution with paramters $\pi_{k,t}^{ i } $ and $\xi_{k,t,l}^{ i }$.
\begin{equation}
    \label{eq:post_margin}
	\begin{aligned}
		&p\left(x_{k, t} \mid \boldsymbol{ \mathcal{Y} }_{\eta, t, :}\right) \propto  p_{0}(x_{k, t}) \mathcal{CN}(C_{ k,t }^{ i }, D_{ k,t }^{ i }) \\
        &=\left(1-\pi_{k, t}^{i}\right) \delta\left(x_{k, t}\right) 
		+\pi_{k, t}^{i} \frac{\sum_{l=1}^{L} \xi_{k, t, l}^{i} \delta\left(x_{k, t}-a_{l}\right)}{\sum_{l=1}^{L} \xi_{k, t, l}^{i}},
	\end{aligned}
\end{equation}
where $\pi_{k,t}^{ i } $ is the sparsity ratio, which is also called the belief indicator. Besides, $\xi_{k,t,l}^{ i } $ is posterior probability that $x_{k,t}$ equals constellation symbol $a_l$ calculated in the $i$-th iteration. The specific expressions of $\pi_{k,t}^{ i } $ and $\xi_{k,t,l}^{ i } $ are as follows

\begin{equation}
	\label{Eq:pi_xi}
	\begin{aligned}
		\xi_{k, t, l}^{i} & =\exp \left[-{\left( \left|a_{l}\right|^{2}-2 \Re \mathfrak{e}\left\{a_{l}^{*} C_{k, t}^{i}\right\}\right) }/{D_{k, t}^{i}}\right] , \\
		\pi_{k, t}^{i} & =\left[1+\left(1-\gamma_{k, t}\right) /\left(\frac{\gamma_{k, t}}{L} \sum_{l=1}^{L} \xi_{k, t, l}^{i}\right)\right]^{-1}. \\
	\end{aligned}
\end{equation}

Therefore, we can derive the posterior mean and variance of transmitted data as follows
\begin{equation}
	\label{Eq:post_mean_var}
	\begin{aligned}
		\hat{x}_{k, t}^{i} & =\left(\pi_{k, t}^{i} \sum_{l=1}^{L} a_{l} \xi_{k, t, l}^{i}\right) /\left(\sum_{l=1}^{L} \xi_{k, t, l}^{i}\right), \\
		v_{k, t}^{i} & =\left(\pi_{k, t}^{i} \sum_{l=1}^{L}\left|a_{l}\right|^{2} \xi_{k, t, l}^{i}\right) /\left(\sum_{l=1}^{L} \xi_{k, t, l}^{i}\right)-\left|\hat{x}_{k, t}^{i}\right|^{2} .
	\end{aligned}
\end{equation}

However, the noise variance $\sigma^2$ and the sparsity ratio $\gamma_{k, t}$ in \text{a prior} distribution is usually unknown in practice. Therefore, we adopt expectation maximization (EM) algorithm to learn these unknown parameters. The update rule of EM algorithm can be written as follows
\begin{equation}
	\label{Eq:EM_Update}
	\begin{aligned}
		\gamma_{k, t}^{i} & =\pi_{k, t}^{i},\\
		(\sigma^2)_{t}^{i} & = \frac{ 1 }{ M } \sum_{ m } \left[\frac{\left|\boldsymbol{\mathcal{Y}}_{\eta, m, t}-Z_{m, t}^{i}\right|^{2}}{\left|1+V_{m, t}^{i} / (\sigma^2)_{ t}^{i-1}\right|^{2}}+\frac{(\sigma^2)_{t}^{i-1} V_{m, t}^{i}}{(\sigma^2)_{t}^{i-1}+V_{m, t}^{i}}\right] .
	\end{aligned}
\end{equation} 
Therefore, the noise variance $\sigma^2$ and the sparsity ratio $\gamma_{ k,t }$ in equation (\ref{Eq:update_factor}) and (\ref{Eq:update_variable}) are replaced by $(\sigma^2)_{t}^{i-1}$ and $\gamma_{ k,t }^{i-1}$ respectively. Besides, we initialize the noise variance $\sigma^{ 2 } $ and sparsity ratio $\gamma_{ k,t } $  as the way in \cite{Vila2013EMAMP}. 

However, the update rule for sparsity $\gamma_{k, t}$ in equation (\ref{Eq:EM_Update}) fails to exploit the structured sparsity of uplink signal. Thus, according to nearest neighbor sparsity pattern learning (NNSPL) in \cite{Meng2016NNSPL}, we update the sparsity ratio as $\gamma_{k, t}^{i}= (1/T) \sum_{t=1}^{T} \pi_{k, t}^{i}$.

Finally, when the number of iterations reaches a predefined threshold, we can estimate the indices of active UEs taking advantage from the belief indicator $\pi_{k,t}^{ i } $. If the average value of the belief indicator with respect to the $k$-th UE is greater than $0.5$, then the $k$-th UE is regarded to be active. Then, we sort the index of UEs regarded to be active from small to large forming the ordered set $\hat{\mathcal{K}}_{ a } $. The number of UEs detected to be active is denoted as $\hat{K}_{ a } = \left| \hat{\mathcal{K}}_{ a }  \right|_{ c }   $. The overall procedure of coarse DD is shown in \textbf{Algorithm \ref{Alg:Coarse_Detection}}.

\begin{algorithm}[t]
	\caption{Coarse Data Detection}
	\label{Alg:Coarse_Detection}
	\begin{algorithmic}[1]
		\REQUIRE Signal received by the beacon antenna of BS $\boldsymbol{ \mathcal{Y} }_{ \eta,:,: }$, spreading sequences ${\bf S}$, damping parameter $\rho_{ \textrm{damp} } $, and the maximum number of iteration $N_{\textrm{coarse} }$.
		\ENSURE Estimation of data $\hat{\mathbf{ X }}$, posterior sparsity ratio $\pi_{k,t}$, $\forall k,t$, indices set of active UEs $\hat{\mathcal{K}}_{ a } $.
		\STATE Initialize $\gamma_{k,t}^{0}$ and $(\sigma^2)_{t}^{0}$ as in \cite{Vila2013EMAMP}.  $V_{m,t}^0 = 1$, $Z_{m,t}^0 = \boldsymbol{ \mathcal{Y} }_{\eta,m,t}$, $\hat x_{k,t}^{0} = 0$, $v_{k,t}^{0} = 1$.
		\label{Step:Initial}
		\FOR{$i=1,2,\cdots,N_{ \textrm{coarse} } $}
		\STATE $\forall m,t$: Update $V_{m,t}^i$ and $Z_{m,t}^i$ according to (\ref{Eq:update_factor}).
		\label{Step:Fac_Update}
		\STATE $V_{m,t}^i = {\rho_{ \textrm{damp} } }{V_{m,t}^{i-1}} + \left(1-\rho_{ \textrm{damp} } \right){V_{m,t}^i}$.
		\label{Step:Damp1}
		\STATE $Z_{m,t}^i = {\rho_{ \textrm{damp} } }{Z_{m,t}^{i-1}} + \left(1-\rho_{ \textrm{damp} } \right){Z_{m,t}^i}$.
		\label{Step:Damp2}
		\STATE $\forall k,t$: Calculate $D_{k,t}^i$ and $C_{k,t}^i$ according to (\ref{Eq:update_variable}).
		\STATE  $\forall k,t,l$: Calculate $\pi_{k,t}^{ i } $ and $\xi_{k,t,l}^{ i } $ according to (\ref{Eq:pi_xi})
		\STATE $\forall k,t:$ Calculate  $\hat{x}_{k,t}^{i}$ and $v_{k,t}^{i}$ according to (\ref{Eq:post_mean_var})
		\label{Step:Var_Update}
		\STATE $\forall t$:  Calculate $(\sigma^2)_{t}^{i}$ according to (\ref{Eq:EM_Update}).
		\STATE $\forall k,t$: Calculate  $\gamma_{k, t}^{i}= (1/T) \sum_{t=1}^{T} \pi_{k, t}^{i}$
		\ENDFOR
		\STATE $\forall k$ if $(1/T) \sum_{t=1}^{T}\pi_{k,t}^{i} > 0.5$, then $\hat\alpha _{ k } = 1$. $\hat{\mathcal{K}}_{ a }=\left\{k|\hat{\alpha} _{ k } = 1 \right\} $
		\label{Step:Active_User}
		\RETURN $\hat{\mathcal{K}}_{ a }$. $\forall k,t$, $[\hat{\mathbf{ X }}]_{ k,t } =\hat{x}_{k,t}^{N_{\textrm{coarse}}}$, $\pi_{k,t}^{N_{\textrm{coarse}}}$
		.
	\end{algorithmic}
\end{algorithm}
\setlength{\textfloatsep}{6pt}

\subsection{The Second Module: Data-Aided Channel Estimation}

We can regard the estimated data symbol $\hat{x}_{ k,t } $ as pilot and perform CE. The model for CE is shown as follows
\begin{equation}
	\label{Eq:CE_matrix}
	 \boldsymbol{ \mathcal{Y} }_{ :,m,: }^{\mathrm{T}} =\tilde{\boldsymbol{\Phi}}_{\mathrm{CE}, m} (\tilde{\boldsymbol{ \mathcal{H} }}_{:,m,:}^{\mathrm{equ}})^\mathrm{T} + \boldsymbol{ \mathcal{W} }_{ :,m,: } ^{\mathrm{T}}  , m=1,2,\cdots,M,
\end{equation}where $\boldsymbol{ \mathcal{Y} }_{ :,m,: }^{\mathrm{T}} \in \mathbb{C}^{T \times N}$ denotes the signal received by the BS with respect to the $m$-th subcarrier, $\tilde{\boldsymbol{\Phi}}_{\mathrm{CE}, m} \in \mathbb{C}^{T \times \hat{K}_{a}}$ refers to the sensing matrix consists of the estimated data symbols. The elements of $\tilde{\boldsymbol{\Phi}}_{\mathrm{CE}, m}$ are ${\left[\tilde{\boldsymbol{\Phi}}_{\mathrm{CE}, m}\right]_{t, \kappa} }  = \left[ \mathbf{ S } \right]_{ m, [\mathcal{\hat{K}}_{ a } ]_{ \kappa }  }    \left[ \hat{\mathbf{ X }}\right] _{\left[\hat{\mathcal{K}}_{a}\right]_{\kappa}, t}$, where $\kappa = 1, 2,\cdots, \hat{K_{ a }} $ is the index of the ordered set $\hat{\mathcal{K}}_{ a }$. $(\tilde{\mathbf{ H }}_{:,m,:}^{\mathrm{equ}})^\mathrm{T} \in \mathbb{C}^{\hat{K}_{a} \times N}$ is the transposed slice of the tensor $\tilde{\boldsymbol{ \mathcal{H} }}^{ \mathrm{equ} } \in \mathbb{C}^{  N \times M \times \hat{K}_{ a }  } $, which represents the small-scale fading factor of the equivalent CSI. The elements of $\tilde{\boldsymbol{\mathcal{H}}}^{ \mathrm{equ} } $ are the product of accurate CSI and pre-equalization factors, i.e., $\tilde{h}^{ \mathrm{equ} }_{ n,m,\kappa }   =h_{n, m,\left[\hat{\mathcal{K}}_{a}\right]_{\kappa}} \theta_{m, \left[\hat{\mathcal{K}}_{ a } \right]_{\kappa} }$. Here, we estimate the small-scale fading factor of equivalent CSI instead of the accurate CSI because the pre-equalization factors are unknown at BS. However, the proportion between channel coefficient of different antennas is not changed by pre-equalization factors. Hence, the equivalent CSI reserves complete information in the virtual angular domain of the accurate CSI, which is still sparse in the virtual angular domain.

Subsequently, we transform the model in equation (\ref{Eq:CE_matrix}) into the virtual angular domain. The transformation is realized by multiplying the conjugation of the Fourier matrix $\mathbf{U}_{\mathrm{BS}} \in \mathbb{C}^{N\times N}$, which is shown as follows
\begin{equation}
	\label{Eq:CE_Angular}
	\mathbf{R}_{m}=\boldsymbol{ \mathcal{Y} }_{:,m,:}^{\mathrm{T}}  \mathbf{U}_{\mathrm{BS}}^{*}=\tilde{\boldsymbol{ \Phi }}_{\mathrm{CE}, m} \tilde{\mathbf{A}}_{m}+\boldsymbol{ \mathcal{W} }^{ \mathrm{A} }_{ :,m,: }  , m=1,2,\cdots M,
\end{equation}
where $\tilde{\mathbf{A}}_{m}=\tilde{\boldsymbol{ \mathcal{H} }}_{:,m,:}^{\mathrm{equ}} \mathbf{U}_{\mathrm{BS}}^{*} \in \mathbb{C}^{\hat{K}_{a} \times N}$ is the sparse equivalent channel matrix in the virtual angular domain on the $m$-th subcarrier. $\boldsymbol{ \mathcal{W} }^{ \mathrm{A} }_{ :,m,: } =  \boldsymbol{ \mathcal{W} }_{:,m,:}^{\mathrm{T}} \mathbf{U}_{\mathrm{BS}}^{*}$ represents the noise in the virtual angular domain. The estimation of $\tilde{\mathbf{A}}_{m}$ can be formulated as a CS problem.

Consequently, we utilize the generalized multiple measurement vector approximate message passing (GMMV-AMP) algorithm in \cite{Ke2020CSADCEMIMO} to solve the CS problem in equation (\ref{Eq:CE_Angular}). Following this process, we estimate the equivalent CSI $\hat{\tilde{\boldsymbol{ \mathcal{H} }}}^{ \mathrm{equ} } \in \mathbb{C}^{ N \times M \times \hat{K}_{ a }  } $ and the noise variance $\hat{\sigma}^{ 2 }$.

  \subsection{The Third Module: Fine Data Detection and Iterations}

After the channel matrix is estimated, we can get a refined estimation of data symbols using the signal received from all antennas at the BS. The signal model for fine DD can be expressed as follows. For $n = 1, 2, \cdots, N$, we can obtain
\begin{equation}
	\label{Eq:DD_separate}
	\boldsymbol{ \mathcal{Y} }_{n,:,:} = \left( \hat{\tilde{\boldsymbol{ \mathcal{H} }}}^{\mathrm{equ}}_{n,:,:} \circ \tilde{\mathbf{S}} \right) \tilde{\mathbf{X}} + \mathbf{W}_{n,:,:} \\
	= \tilde{\mathbf{\Phi}}_{\mathrm{DD},n} \tilde{\mathbf{X}}  + \boldsymbol{ \mathcal{W} }_{ n,:,: }  ,
\end{equation}
where  $\hat{\tilde{\boldsymbol{ \mathcal{H} }}}^{\mathrm{equ}}_{n,:,:} \in \mathbb{C}^{M\times \hat{K}_{a}}$ denotes the estimation of equivalent CSI with respect to the $n$-th antenna obtained from the data-aided CE module. $\tilde{\mathbf{ S }} = [\mathbf{ S }]_{ :,_{ \hat{\mathcal{K}}_{ a }  }  } \in\mathbb{C}^{ M \times \hat{K_{ a } } } $ and $\tilde{\mathbf{X}}\in \mathbb{C}^{\hat{K}_{ a }  \times T}$ denotes the spreading codes and data symbols sent by UEs which are detected to be active, respectively. Besides, $\tilde{\boldsymbol{ \Phi }}_{ \mathrm{DD},n } =  \hat{\tilde{\boldsymbol{ \mathcal{H} }}}^{\mathrm{equ}}_{n,:,:} \circ \tilde{\mathbf{S}} \in \mathbb{C}^{ M \times \hat{K}_{ a } }$.
 
By stacking the received signals of all antennas, we can obtain
\begin{equation}
	\label{Eq:DD}
	\mathbf{ Y }_{\mathrm{DD}} = \tilde{\boldsymbol{\Phi}}_{\mathrm{DD}}\tilde{\mathbf{X}}+\mathbf{W}_{ \mathrm{DD} },
\end{equation} 
where $\tilde{\boldsymbol{ \Phi }}_{\mathrm{DD}} = [\tilde{\boldsymbol{\Phi}}_{\mathrm{DD},1}^{\mathrm{T}} ,\tilde{\boldsymbol{\Phi}}_{\mathrm{DD},2}^{\mathrm{T}} ,\cdots,\tilde{\boldsymbol{\Phi}}_{\mathrm{DD},N }^{\mathrm{T}} ]^{\mathrm{T}}  \in \mathbb{C}^{MN \times \hat{K}_{a}}$, $\mathbf{ Y }_{\mathrm{DD}} = [\boldsymbol{ \mathcal{Y} }_{1,:,:}^{\mathrm{T}} ,\boldsymbol{ \mathcal{Y} }_{2,:,:}^{\mathrm{T}} ,\cdots,\boldsymbol{ \mathcal{Y} }_{N,:,:}^{\mathrm{T}} ]^{\mathrm{T}} \in \mathbb{C}^{ MN \times T  } $, and $\mathbf{ W }_{ \mathrm{DD} } = \left[ \boldsymbol{\mathcal{W}}_{ 1,:,: }^{\mathrm{T}} , \boldsymbol{\mathcal{W}}_{ 2,:,: }^{\mathrm{T}} , \cdots, \boldsymbol{\mathcal{W}}_{ N,:,: }^{\mathrm{T}} \right]^{\mathrm{T}}  \in \mathbb{C}^{ MN \times T }  $.

We estimate $\tilde{\mathbf{ X }}$ using the linear minimum mean square error (LMMSE) estimator \cite{Ying2023MIMOLEO} as follows
\begin{equation}
	\label{Eq:LMMSE}
	[\hat{\mathbf{X}}]_{\hat{\mathcal{K}}_a,:} \leftarrow \hat{\tilde{\mathbf{X}}} = (\tilde{\boldsymbol{\Phi}}_{\mathrm{DD}}^{\mathrm{H}} \tilde{\boldsymbol{\Phi}}_{\mathrm{DD}} + \hat{\sigma}^{2}\mathbf{I}_{\hat{K}_{a} \times \hat{K}_{ a } })^{-1}\tilde{\boldsymbol{\Phi}}_{\mathrm{DD}}^{\mathrm{H}} \mathbf{ Y }_{\mathrm{DD}},
\end{equation}
where $\leftarrow$ refers to the operation that replaces the value of elements on the left side with the value of elements on the right side. Then, the updated $\hat{\mathbf{ X }}$ is passed to the data-aided CE stage and one iteration is completed. The proposed algorithm iteratively performs the data-aided CE module and fine DD module until the numbers of iterations reaches the predefined threshold $N_{ \textrm{iter} } $. The overall procedure is shown in \textbf{Algorithm \ref{Alg:prop}}.

\begin{algorithm}[t]
	\caption{The Proposed Iterative Detection Scheme}\label{Alg:prop}
	\begin{algorithmic}[1]
		\REQUIRE  Received signal $\boldsymbol{\mathcal{Y}} $, number of iterations $N_{ \textrm{iter} } $.
		\ENSURE  The indices set of active UEs $\hat{\mathcal{K}}_{ a } $, the estimated equivalent CSI $\hat{\tilde{\boldsymbol{\mathcal{H}}}}^{ \mathrm{equ} }$, and the estimated data $\hat{\mathbf{ X }}$.
		\STATE Obtain $\hat{\mathcal{K}}_{ a } $ and $\hat{\mathbf{ X }}$ using \textbf{Algorithm} \ref{Alg:Coarse_Detection}.
		\FOR{$i=1,2,\cdots,N_{ \textrm{iter} } $}
		\STATE Apply the GMMV-AMP algorithm to obtain $\hat{\tilde{\boldsymbol{\mathcal{H}}}}^{ \mathrm{equ} }$ and the noise variance $\hat{\sigma}^{ 2 }$ using $\hat{\mathbf{ X }}$.
		\STATE Perform fine DD  to obtain the estimated data of active UEs $\hat{\tilde{\mathbf{ X }}}$ according to (\ref{Eq:LMMSE}) using $\hat{\tilde{\boldsymbol{\mathcal{H}}}}^{ \mathrm{equ} }$.
		\STATE Update the estimated data as $[\hat{\mathbf{ X }}]_{ \hat{K}_a,: } \leftarrow \hat{\tilde{\mathbf{X}}}$.
		\ENDFOR
		\RETURN $\hat{\mathcal{K}}_a$, $\hat{\mathbf{ X }}$, $\hat{\tilde{\boldsymbol{\mathcal{H}}}}^{ \mathrm{equ} }$. 
	\end{algorithmic}
	\label{alg1}
\end{algorithm}

\section{Simulation Results}

We consider that the BS employs a ULA with $N = 128$ antennas. The total number of UEs is set as $K=500$, and $K_{a} = 50$ UEs are active in each frame. The threshold for pre-equalization is set to $h_{0} = 0.2$. The modulation order is set as $L=4$. The element of spreading codes $\mathbf{ s }_{ k } $ is generated following independent and identically distributed (i.i.d.) complex Gaussian distribution $\mathcal{CN}\left( 0,1 \right)$. The transmit power of all UEs are equally set as $\rho=7\mathrm{dBm}$. The parameters of wireless channel are set as follows. We consider the number of paths for each UE is a random integer variable that obeys uniform distribution from $8$ to $12$. The angle spreading is set to $\Delta = 7.5^{\circ}$. The large scale fading follows the Log-distance path loss model as $g_{ k }=128.1+37.6\log_{ 10 }\left( d_{ k }  \right)$ with the distance $ d_{ k } $ measured in $\mathrm{km}$, which obeys the uniform distribution $d_{ k } \sim \mathcal{U}\left[ 0.1\mathrm{km}, 1\mathrm{km} \right]$. The power spectrum density of noise is $-174\mathrm{dBm/Hz}$ and the system bandwidth is set as $B_{ s } =  10\mathrm{MHz}$.  The parameters mentioned above are fixed in the following simulations unless otherwise specified.

We use the activity detection error probability (ADEP), and bit error rate (BER) in \cite{Mei2022Preeq} as the metrics of AD and DD, respectively. Besides, we use the normalized mean square error (NMSE) between the estimated small fading factor of equivalent CSI and the true small fading factor of equivalent CSI to measure the CE performance.

\begin{table}
\centering
\footnotesize
\setlength{\tabcolsep}{3pt}
\renewcommand{\arraystretch}{1.3}
\begin{tabular}{m{1.3cm}|m{1.6cm}|m{1.5cm}|m{1.8cm}|m{1.5cm}}
\hline
\textbf{Methods} & \textbf{Uplink schemes} & \textbf{Detection problems} & \textbf{Number of BS antennas} & \textbf{Reconstruction algorithms} \\
\hline
Baseline 1 & preamble+data & JADCE \& coherent DD & multiple antennas& GAMP-MMV\&+LMMSE\\
\hline
Baseline 2 & preamble+data & JADCE \& coherent DD & multiple antennas& SOMP+LMMSE \\
\hline
Baseline 3 & \shortstack{pre-eq. aided} & JADD & single antenna& OAMP \\
\hline
Baseline 4 & \shortstack{pre-eq. aided} & JADD & single antenna& SOMP \\
\hline
Proposed & \shortstack{pre-eq. aided} &JADD+CE\& +refined DD & multiple antennas& AMP+LMMSE \\
\hline
\end{tabular}
\caption{\protect Comparison of methods and their characteristics}
\label{tab:algorithm_comparison}
\end{table}

The following  four grant-free massive access baseline methods shown in table \ref{tab:algorithm_comparison} are considered to verify the superiority of the proposed scheme. 
\textbf{Baseline 1:}  The mMIMO-based grant-free massive access scheme that adopts the ``pilot+data" frame structure which utilizes GMMV-AMP algorithm \cite{Ke2020CSADCEMIMO} for JADCE and adopts LMMSE for coherent DD. \textbf{Baseline 3:} Same as Baseline 1 except that JADCE is performed with simultaneous orthogonal matching pursuit (SOMP) algorithm.\cite{Determe2017SOMP} For the fairness of comparison, we set the pilot sent in Baseline 1 and Baseline 3 to have the same uplink access latency and occupies the same frequency resources as the proposed scheme. Specifically, in Baseline 1, we set the number of subcarriers and the pilot time slot overhead to be $M$ and $T$, respectively.
\textbf{Baseline 3:} The pre-equalization aided grant-free massive access scheme tailored for single-antenna BS configurations proposed in \cite{Mei2022Preeq}. \textbf{Baseline 4:} Compared with Baseline 3, the difference is that the JADD is conducted with SOMP.  The CSI is assumed to be accurately estimated at UEs' side for pre-equalization so that the BS is capable of performing AD and DD without uplink pilot and CE. Hence, only the ADEP and BER performance of Baseline 3 and Baseline 4 is investigated in simulations. To ensure the consistency and the fairness of comparison, the spreading codes of Baseline 3 are randomly drawn from Fourier matrix as that in \cite{Mei2022Preeq}. Besides, the pre-equalization procedure in equation (\ref{Eq:Pre-eq}) is considered in Baseline 3.

\begin{figure}[t]
    \centering
    \subfloat[BER performance]{
    \label{fig:proposed_ber}
    \includegraphics[width=0.23\textwidth]{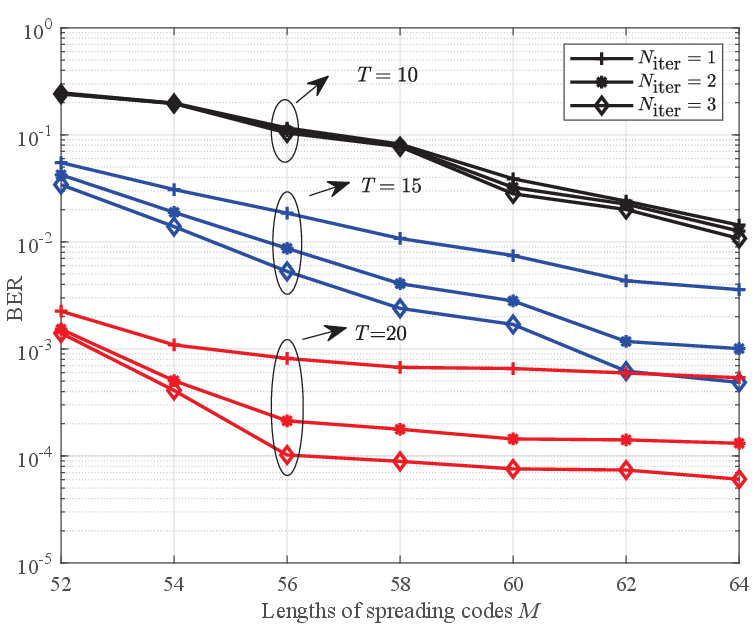}}
    \subfloat[NMSE performance]{
    \label{fig:proposed_nmse}
    \includegraphics[width=0.23\textwidth]{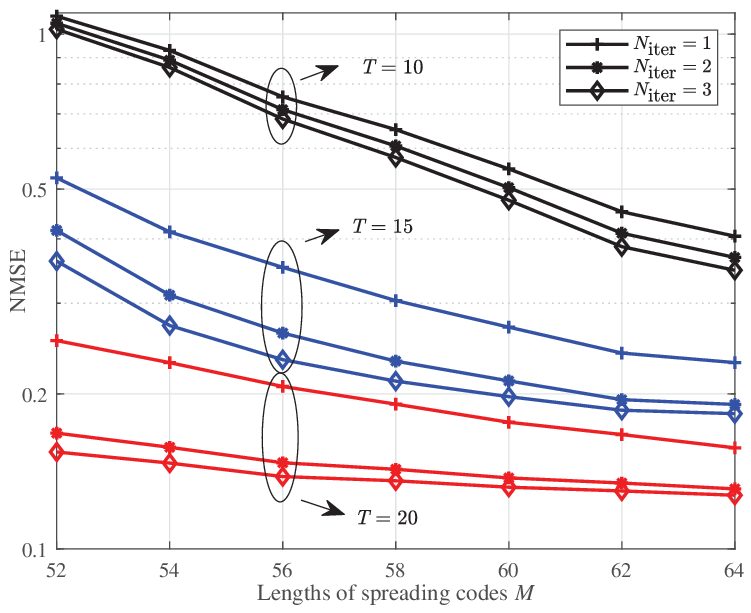}}
    \caption{The BER and NMSE performance of different schemes with transmit power $\rho = 7\textrm{dBm}$.}
    \label{fig:proposed}
\end{figure}

Fig. \ref{fig:proposed} displays the BER and NMSE performance of the proposed scheme under different time slot overhead $T$, lengths of spreading codes $M$, and the numbers of iterations $N_{ \textrm{iter} } $. Here, only the curves with  $N_{ \textrm{iter} } $ from 1 to 3 are shown, since the performance of the proposed scheme has no further improvement when $N_{ \textrm{iter} }>3 $. From Fig. \ref{fig:proposed}, it can be concluded that the BER and NMSE performance both get enhanced with the increase of  $N_{ \textrm{iter} } $, which showcases the effectiveness of the iteration procedure. 
Besides, the BER and NMSE performance gets better when $M$ and $T$ increase. The reason is that the measurement of indeterminacy for CS problems in coarse DD and data-aided CE are $M/K$ and $T/\hat{K}_a$, respectively. As $M$ and $T$ increase, the indeterminacy in both the coarse DD and data-aided CE problems decreases. This reduction in uncertainty improves the accuracy of recovering sparse signals.

\begin{figure}[t]
    \centering
    \begin{minipage}[b]{0.23\textwidth}
        \centering
        \includegraphics[width=\textwidth]{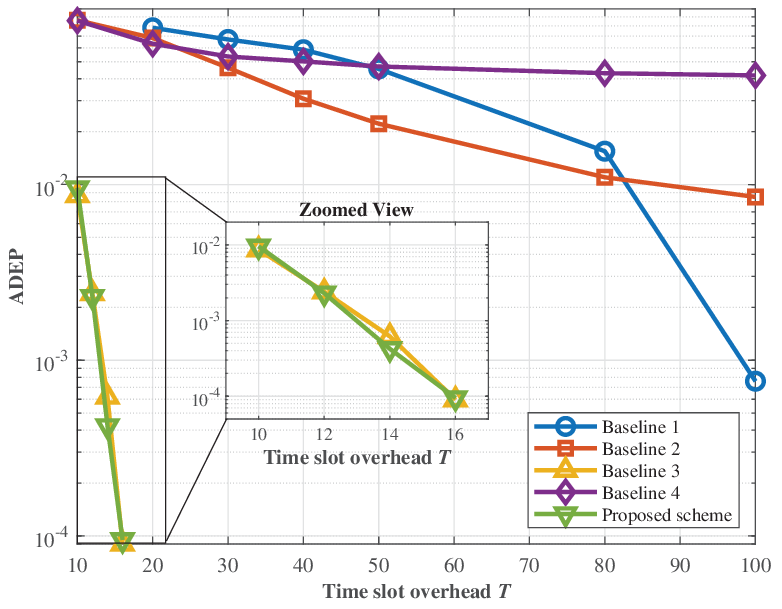}
        \caption{Comparison of ADEP performance for different schemes versus time slot overhead $T$ with $M=56, \rho = 7 \textrm{dBm}$.}
        \label{fig:adep_t}
    \end{minipage}
    \hspace{3pt}
    \begin{minipage}[b]{0.23\textwidth}
        \centering
        \includegraphics[width=\textwidth]{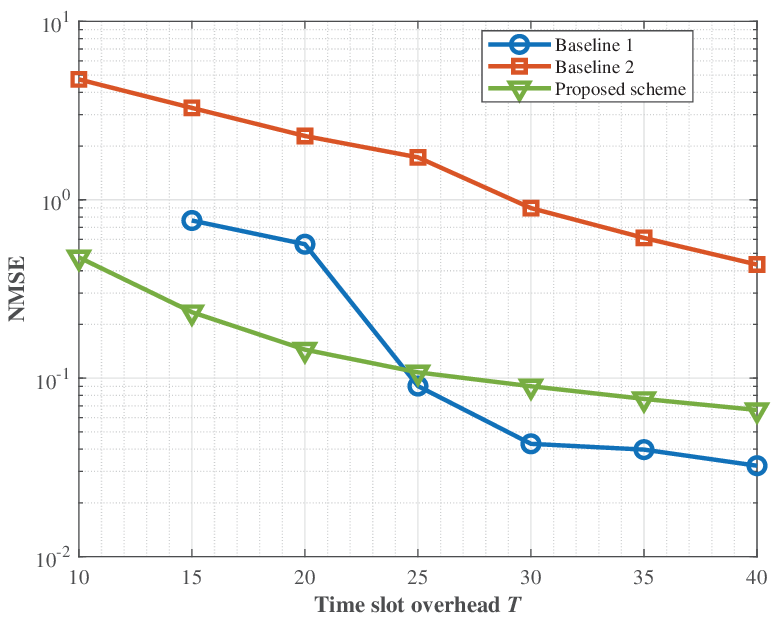}
        \caption{Comparison of NMSE performance for different schemes versus time slot overhead $T$  with $M=60, \rho = 7 \textrm{dBm}$.}
        \label{fig:nmse_t}
    \end{minipage}
    \label{fig:adep_nmse}
\end{figure}

\begin{figure}[t]
    \centering
    \subfloat[BER performance versus time slot overheads $T$  with $\rho=20$.]{
    \label{fig:ber_t}
    \includegraphics[width=0.23\textwidth]{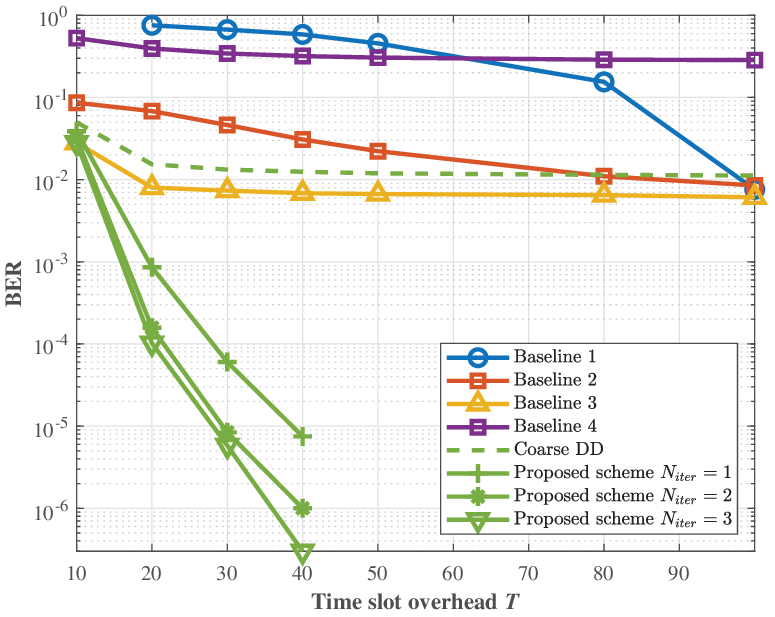}}
    \hspace{3pt}
    \subfloat[BER performance versus transmit power $\rho$  with $T=20$.]{
    \label{fig:ber_p}
    \includegraphics[width=0.23\textwidth]{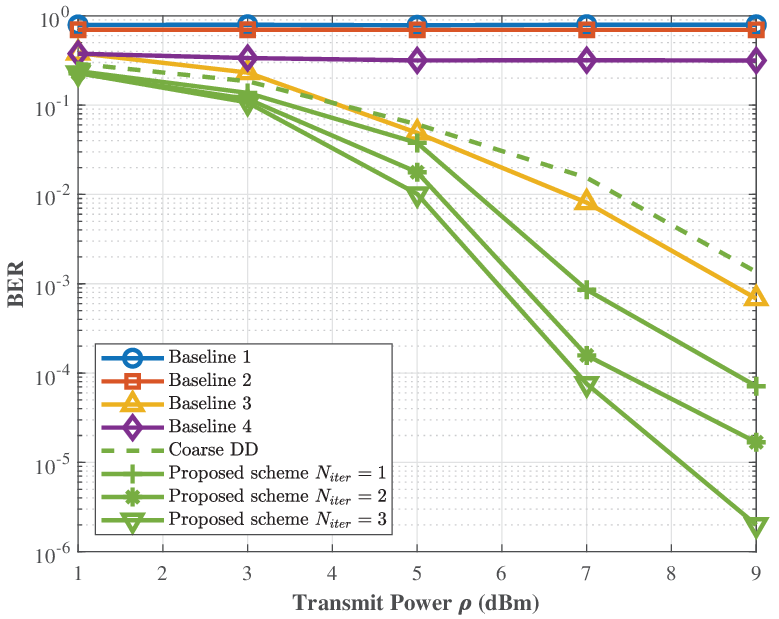}}
    \caption{The BER performance of different schemes with $M=60$.}
    \label{fig:ber}
\end{figure}

Fig. \ref{fig:adep_t} shows the performance of AD versus the time slot overhead $T$, where the length of spreading codes is set to $M=56$ and the number of iterations is set to $ N_{ \textrm{iter} } = 3 $. For the proposed scheme and baseline method that adopts 'pilot+data' frame structure (Baseline 1 and 2), the time slot overhead refers to the number of OFDM symbols within one frame $T$ and the time slot overhead of the pilot signal, respectively. As evidenced by Fig. \ref{fig:adep_t}, the proposed scheme outperforms other baselines. This is due to the AD of Baseline 1 and Baseline 2 necessitates the accurate estimation of the high-dimensional mMIMO channel matrix, which requires a substantial number of pilot symbols. Baseline 2 and Baseline 4 adopts greedy-based CS algorithms that cannot exploit the prior information of the signal, thus have relatively lower AD performance than the proposed scheme. Moreover, the \textit{a priori} distributions of data symbols in the proposed scheme and Baseline 3 are accurate, which ensures the good AD performance. However, the i.i.d. Bernoulli-Gaussian \textit{a priori} distribution of CSI in Baseline 1 failed to model the coherence between antennas, which affects the performance of AD. Apart from that, we can conclude from Fig. \ref{fig:adep_t} that the AD performance of the proposed scheme is almost the same as Baseline 1 with the single-antenna BS configuration. The reason for that is the coarse DD step for AD does not leverage the diversity of multiple antennas. 

Fig. \ref{fig:nmse_t} compares the CE performance of different schemes, where the length of spreading codes is set to $M=60$ and the number of iterations is set to $N_{ \textrm{iter} } = 3$. With a slight abuse of notations, we use $T$ to represent the time slot overhead of each scheme. For $T < 25$, The proposed scheme outperforms Baseline 1 and Baseline 2, while surpassed by Baseline 1 when $T \geq 25$. The reason for this phenomenon is twofold. First, in Baseline 1, the pilot signal is known at BS. However, in the proposed scheme, the BS uses the estimated data in coarse DD as pilot, which can be regraded as a semi-blind estimation that has a relatively degraded performance. Second, the measure of indeterminacy of the CS problems for Baseline 1 and the proposed algorithm is $T/K$ and $T/\hat{K}_{ a } $, respectively. Hence, the indeterminacy for Baseline 1 is higher than the proposed scheme, which decreases Baseline 1's probability for accurate recovery of the signal. This effect is exacerbated when the dimension of observations $T$ is small. For example, when $T = 10$, Baseline 1 failed to converge due to the high indeterminacy. Hence, the proposed scheme is more suitable for the case that the access latency is required to be very low.

Fig. \ref{fig:ber_t} shows the BER performance of different schemes versus time slot overhead $T$ given $M=60$. Fig. \ref{fig:ber_p} shows the BER performance versus UEs' transmit power $\rho$ with $M=60$, $T=20$. The green lines with legend `Coarse DD' refers to the performance of coarse DD module in \textbf{Algorithm \ref{Alg:Coarse_Detection}}. Other green lines refers to the performance of the proposed scheme with number of iterations shown in the legend. In Fig. \ref{fig:ber_t}, the performance of the proposed algorithm with $T \geq 50$ is not shown. This is because when $T \geq \hat{K_{ a }} $, the problem in equation (\ref{Eq:CE_matrix}) turns into deterministic or overdetermined problem, where the GMMV-AMP cannot converge. Note that, under the case, we can resolve the CE problem with LMMSE estimation. It can be seen from both figures that the BER performance of Baseline 1 and Baseline 2 is relatively poor when the number of pilot is small. This is due to its inaccurate activity detection. The proposed scheme can achieve the same BER performance as Baseline 1 with much lower access latency. Besides, the BER of the proposed scheme is superior to that of Baseline 3, Baseline 4 and coarse DD, which is due to the diversity of multiple antennas.

\section{Conclusion}

This paper proposes a pre-equalization aided grant-free massive access scheme for mMIMO systems. Specifically, the BS first broadcasts a beacon signal with one activated beacon antenna that enables UEs to perform uplink transmission with pre-equalization associated with the channel of the beacon antenna. The pre-equalization transforms the blind detection problem that both channel and data are unknown into three classic linear models, i.e., the coarse DD module, the data-aided CE module, and the fine DD module. Specifically, the coarse DD module performs JADD without the knowledge of CSI, which yields the detected active UEs and a coarse estimation of data symbols. Then, the algorithm iteratively performs data-aided CE and fine DD. Particularly, leveraging the sparsity of uplink signal in multiple domains, we model the coarse DD and data-aided CE as CS sparse signal recovery problems and resolve it using the AMP algorithm. Finally, the simulation results demonstrate the efficacy of the proposed iterative detection scheme, where the DD performance of the proposed scheme outperforms state-of the-art grant-free massive access schemes under the same access latency.


\begin{thebibliography}{1}
\bibliographystyle{IEEEtran}

\bibitem{Gao2024CSGFMA}
Z. Gao \textit{et al.}, ``Compressive-sensing-based grant-free massive access for 6G massive communication," \textit{IEEE Internet Things J.}, vol. 11, no. 5, pp. 7411-7435, March 2024.

\bibitem{Chen2021MA5G}
X. Chen \textit{et al.}, ``Massive Access for 5G and Beyond," \textit{IEEE J. Sel. Areas Commun.}, vol. 39, no. 3, pp. 615-637, March 2021.

\bibitem{Liu2018massiveconnectivityMIMO}
L. Liu \textit{et al.}, ``Massive connectivity with massive MIMO—part I: device activity detection and channel estimation," \textit{IEEE Trans. Signal Process.}, vol. 66, no. 11, pp. 2933-2946, June 2018.

\bibitem{Ke2020CSADCEMIMO}
M. Ke \textit{et al.}, ``Compressive sensing-based adaptive active UE detection and channel estimation: Massive access meets massive MIMO," \textit{IEEE Trans. Signal Process.}, vol. 68, pp. 764-779, Jan. 2020.

\bibitem{Ke2021cellfree}
M. Ke \textit{et al.}, ``Massive access in cell-free massive MIMO-based internet of things: Cloud computing and edge computing paradigms," \textit{IEEE J. Sel. Areas Commun.}, vol. 39, no. 3, pp. 756-772, March 2021.

\bibitem{Chen2022JADCE}
W. Chen \textit{et al.}, ``Joint activity detection and channel estimation in massive MIMO systems with angular domain enhancement," \textit{IEEE Trans. Signal Process.}, vol. 21, no. 5, pp. 2999-3011, May 2022.

\bibitem{Zhou2023NOMAOTFS}
X. Zhou \textit{et al.}, ``Active terminal identification, channel estimation, and signal detection for grant-Free NOMA-OTFS in LEO satellite Internet-of-Things," \textit{IEEE Trans. Wireless Commun.}, vol. 22, no. 4, pp. 2847-2866, April 2023.

\bibitem{Zheng2024DLADCEIRS}
S. Zheng \textit{et al.}, ``Hybrid driven learning for joint activity detection and channel estimation in IRS-assisted massive connectivity," \textit{IEEE Trans. Signal Process.}, early access, March 19, 2024.

\bibitem{Shen2022MIMOOTFS}
B. Shen \textit{et al.}, ``Random access with massive MIMO-OTFS in LEO satellite communications,” \textit{IEEE J. Select. Areas Commun.}, vol. 40, no. 10, pp. 2865–2881, 2022.

\bibitem{Shao2022lowrank}
X. Shao \textit{et al.}, ``Exploiting simultaneous low-rank and sparsity in delay-angular domain for millimeter-wave/terahertz wideband massive access,” \textit{IEEE Trans. Wireless Commun.}, vol. 21, no. 4, pp. 2336–2351, 2022.

\bibitem{Ge2021OTFSNOMA}
Y. Ge \textit{et al.}, ``OTFS signaling for uplink NOMA of heterogeneous mobility users,” \textit{IEEE Trans. Wireless Commun.}, vol. 69, no. 5, pp. 3147–3161, 2021.

\bibitem{Wei2017JADD}
C. Wei, H. Liu, Z. Zhang, J. Dang and L. Wu, ``Approximate message passing-based joint sser activity and data detection for NOMA," in IEEE Communications Letters, vol. 21, no. 3, pp. 640-643, March 2017.

\bibitem{Mei2022Preeq}
Y. Mei \textit{et al.}, ``Compressive sensing-based joint activity and data detection for grant-free massive IoT access," \textit{IEEE Trans. Wireless Commun.}, vol. 21, no. 3, pp. 1851-1869, March 2022.

\bibitem{Gao2015SparsityMIMO}
Z. Gao \textit{et al.}, ``Spatially common sparsity based adaptive channel estimation and feedback for FDD massive MIMO," \textit{IEEE Trans. Signal Process.}, vol. 63, no. 23, pp. 6169-6183, Dec. 2015.  

\bibitem{Vila2013EMAMP}
J. P. Vila \textit{et al.}, ``Expectation-Maximization Gaussian-mixture approximate message passing," \textit{IEEE Trans. Signal Process.}, vol. 61, no. 19, pp. 4658-4672, Oct 2013.

\bibitem{Meng2016NNSPL}
X. Meng \textit{et al.}, ``Approximate message passing with nearest neighbor sparsity pattern learning,” 2016, \textit{arXiv:1601.00543}.

\bibitem{Donoho2010AMP}
D. L. Donoho \textit{et al.},``Message passing algorithms for compressed sensing: I. motivation and construction,” in 2010 IEEE Information Theory Workshop on Information Theory (ITW 2010, Cairo), 2010, pp. 1–5.

\bibitem{Determe2017SOMP}
J. -F. Determe \textit{et al.}, ``On the noise robustness of simultaneous orthogonal matching pursuit," \textit{IEEE Trans. Signal Process.}, vol. 65, no. 4, pp. 864-875, 15 Feb.1, 2017.

\bibitem{Ying2023MIMOLEO}
K. Ying \textit{et al.}, ``Quasi-synchronous random access for massive MIMO-based LEO satellite constellations," \textit{IEEE J. Select. Areas Commun.}, vol. 41, no. 6, pp. 1702-1722, June 2023.

\end{thebibliography}

\end{document}